\documentclass[journal]{IEEEtran}

\usepackage{color}%
\usepackage[cmex10]{amsmath}
\usepackage{amssymb}
\usepackage{epsfig}
\usepackage{subfigure}
\usepackage{stfloats}
\usepackage{cite}
\usepackage{enumerate}
\usepackage{graphicx}
\usepackage{amsmath}
\usepackage{bm}
\usepackage{amsmath}
\usepackage{mathtools}
\usepackage{multicol}
\usepackage{yhmath}
\usepackage{polynom}
\usepackage{booktabs}
\usepackage{float}



\begin{document}
\title{Security Enhancement for Coupled Phase-Shift STAR-RIS Networks}
\author{Zheng Zhang, Zhaolin Wang, Yuanwei Liu, Bingtao He, Lu Lv, and Jian Chen
\thanks{Z. Zhang, B. He, L. Lv, and J. Chen are with the State Key Laboratory of Integrated Services Networks, Xidian University, Xi'an 710071, China (e-mail: zzhang\_688@stu.xidian.edu.cn; bthe@xidian.edu.cn; lulv@xidian.edu.cn; jianchen@mail.xidian.edu.cn).}
\thanks{Z. Wang and Y. Liu are with the School of Electronic Engineering and Computer Science,
Queen Mary University of London, London E1 4NS, U.K. (e-mail: zhaolin.wang@qmul.ac.uk; yuanwei.liu@qmul.ac.uk).}
}
\maketitle

\begin{abstract}
    The secure transmission of the simultaneously transmitting and reflecting reconfigurable intelligent surface (STAR-RIS) aided communication system is investigated. Considering the \textit{coupled} phase shifts of STAR-RISs and the fair secrecy requirement of users, a novel secure beamforming design is proposed for addressing the unique full-space mutual eavesdropping of STAR-RIS aided communication. In particular, a penalty based secrecy beamforming algorithm is developed to solve the resulting non-convex optimization problem, where the closed-form solutions of the coupled transmission/reflection coefficients are obtained in each iteration. Numerical results demonstrate that 1) the proposed scheme achieves higher secrecy capacity than conventional RIS; 2) 4-bit discrete phase shifters are sufficient for secrecy guarantee.
\end{abstract}

\IEEEpeerreviewmaketitle

\section{Introduction}\label{1:int}

To cope with the rapidly growing demands for various emerging applications and ubiquitous wireless communications, reconfigurable intelligent surface (RIS) has been envisioned as a promising key enabler for next generation wireless networks \cite{M.Di.Renzo_RIS_JSAC,Yuanwei_IRS_magazine,Lu_IRS_magazine}. However, the inherent limitation of \emph{half-space} coverage of conventional reflecting-only RIS restricts its application flexibility and electromagnetic propagation adjustment capacity. Against this background, a new principle of simultaneous transmitting and reflecting RIS (STAR-RIS) has been proposed \cite{STAR_magazine}. By splitting the incident signal into transmitted and reflected signals on both sides of the surfaces, STAR-RISs are capable of enabling a \emph{full-space} smart radio environment \cite{J.Xu_STAR-RIS,S.Yang_STAR-RIS,X.Mu_STAR-RIS,H.Niu_STAR-RIS_1}.

However, the unique ability of STAR-RIS to reconfigure full-space radio propagation environments inevitably results in full-space wiretapping. In other words, the eavesdroppers on any side of STAR-RIS can access the confidential information passing through STAR-RIS, which raises more stringent security challenges from the information-theory perspective. Fortunately, it has been claimed that physical layer security (PLS) techniques are expected to secure STAR-RIS aided communications by exploiting intrinsic features of wireless channels, such as fading, noise, and interference \cite{H.Niu_STAR-RIS_security,W.Wang_STAR-RIS_PLS,Y.Han_STAR-RIS_PLS,ZZ_STAR}.


Noteworthy, the security design proposed in the aforementioned works \cite{H.Niu_STAR-RIS_security,W.Wang_STAR-RIS_PLS,Y.Han_STAR-RIS_PLS,ZZ_STAR}, is based on the ideal STAR-RIS model with \textit{independent} phase shifts, which requires complicated semi-passive or active STAR-RIS metasurface architectures. While for low-cost passive lossless STAR-RISs, the recent works  \cite{Y.Liu_coupled_STAR,H.Niu_coupled_STAR} has pointed out that the transmission and reflection phase shifts are \textit{coupled} with each other, which implies there exists a fixed phase-shift difference ($\frac{\pi}{2}$ or $\frac{3\pi}{2}$) between the transmission and reflection coefficients. As such, the existing security design \cite{H.Niu_STAR-RIS_security,W.Wang_STAR-RIS_PLS,Y.Han_STAR-RIS_PLS,ZZ_STAR} cannot ensure the network security as well as the phase-shift coupling constraint, which requires the redesign of the transmit beamforming and STAR-RIS coefficients. Additionally, the conventional sum secrecy capacity maximization schemes in \cite{H.Niu_STAR-RIS_security,W.Wang_STAR-RIS_PLS}, where more transmit power is concentrated on the users with better channel conditions, cannot guarantee fairness among legitimate users in terms of security transmission.




Motivated by the above, this paper studies secure beamforming optimization with the aim of minimum secrecy capacity maximization for a coupled STAR-RIS network. The main contributions of this paper are summarized below.

\begin{itemize}
\item We focus on a coupled phase-shift STAR-RIS aided secrecy communication framework, where a STAR-RIS is deployed on the building facades to transmit and/or reflect the confidential information to the indoor user (IU) and outdoor user (OU) in the presence of multiple eavesdroppers. Based on this framework, a joint optimization problem of transmit beamforming and coupled phase-shift transmission/reflection coefficients is formulated to maximize the minimum secrecy capacity of IU and OU for fairness-security tradeoff consideration.
\item A penalty based secrecy beamforming (PSB) algorithm is proposed to solve the formulated non-convex optimization problem, where an augmented lagrangian (AL) problem is constructed to relax the coupled phase-shift constraints in the outer loop. In the inner loop, the semidefinite relaxation (SDR) technique is employed to obtain the rank-one transmit beamforming, while the first-order optimality condition is adopted to derive the optimal coupled transmission/reflection coefficients.
\item Numerical results demonstrate the convergence of the PSB algorithm. It is also verified that: 1) the proposed secure beamforming scheme is superior to the conventional RIS regarding secrecy performance; and 2) regardless of whether the phase shifts are coupled, quantization of 4 bits is sufficient for discrete phase-shift STAR-RIS to achieve the comparable performance as the case of continuous phase shifts.
\end{itemize}


\section{System Model and Problem Formulation}
\label{sec:format}
\subsection{Network Description}
\begin{figure}[t]
  \centering
  \includegraphics[scale = 0.3]{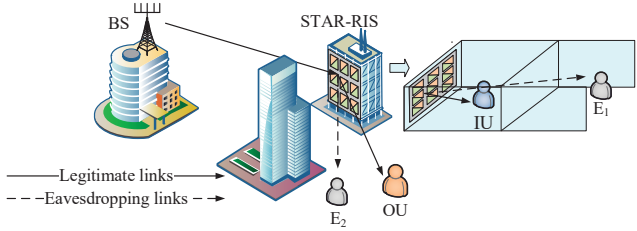}
  \caption{A coupled phase-shift STAR-RIS aided secrecy downlink network.} \vspace{-5mm}
  \label{Fig.1}
\end{figure}

We consider a STAR-RIS aided downlink communication network as shown in Fig. \ref{Fig.1}, in which an $M$-antenna BS exploits an $N$-element STAR-RIS to transmit the confidential signals to a single-antenna IU and a single-antenna OU in presence of two single-antenna eavesdroppers ($\text{E}_{1}$ and
$\text{E}_{2}$). Without loss of generality, we assume that IU is on the transmission side of the STAR-RIS and OU is on the reflection side. $\text{E}_{1}$ and $\text{E}_{2}$ are assumed to be situated near the IU and OU for wiretapping signals of both IU and OU. Due to the obstacles, there is no direct link between the BS and IU/OU/$\text{E}_{1}$/$\text{E}_{2}$. The baseband equivalent channels from STAR-RIS to BS, IU, OU, $\text{E}_{1}$ and $\text{E}_{2}$ are represented as $\mathbf{G}\in\mathbb{C}^{N\times M}$, $\mathbf{h}_{\text{I},\text{S}}\in\mathbb{C}^{N\times 1}$, $\mathbf{h}_{\text{O},\text{S}}\in\mathbb{C}^{N\times 1}$, $\mathbf{h}_{\text{E}_{1},\text{S}}\in\mathbb{C}^{N\times 1}$ and $\mathbf{h}_{\text{E}_{2},\text{S}}\in\mathbb{C}^{N\times 1}$. Since the geographical attributes of STAR-RIS can be delicately selected to favor line-of-sight (LoS) transmission, we adopt the Rician fading model for all the channels. Additionally, to reveal the fundamental secrecy performance limit of the coupled phase-shift STAR-RIS aided wireless communication, we assume that $\text{E}_{1}$ and $\text{E}_{2}$ are other unscheduled users of legitimate network with being trusted at service level but untrusted in data level, such that the full channel state information (CSI) of all the channels are available at the BS \cite{H.Niu_STAR-RIS_security,W.Wang_STAR-RIS_PLS,Y.Han_STAR-RIS_PLS,ZZ_STAR}.

\subsection{Coupled Phase-Shift Model of STAR-RIS}
To simultaneously serve the indoor and outdoor cellular users, a coupled phase-shift STAR-RIS is employed to operate in the energy splitting (ES) mode, where the incident signal energy upon each element is split into the transmitted and reflected parts by a set of complex coefficients, i.e., $\sqrt{\beta^{\text{t}}_{n}}e^{j\theta^{\text{t}}_{n}}$ and $\sqrt{\beta^{\text{r}}_{n}}e^{j\theta^{\text{r}}_{n}}$ ($1\leq n\leq N$). More precisely, the real-valued coefficients $\beta^{\text{t}}_{n},\beta^{\text{r}}_{n}\in[0,1]$ denote the transmission and reflection amplitudes of the \textit{n}-th element of STAR-RIS, and $\theta^{\text{t}}_{n},\theta^{\text{r}}_{n}\in[0,2\pi)$ denote the transmission and reflection phase-shift adjustments. Considering the purely passive lossless hardware architecture of STAR-RIS, the elements are only excited by the incident signals, which produce the magnetization and electric polarization currents and radiate the corresponding transmission and reflection fields. Notably, the excited magnetic and electric currents are required to satisfy the the law of energy conservation and the boundary condition, which indicates that there is an inevitable coupling between transmission and reflection coefficients. Therefore, according to the conservation of energy, the amplitude coefficients should satisfy
\begin{equation}
\label{1}
\beta^{\text{t}}_{n}=1-\beta^{\text{r}}_{n}, \ 1\leq n\leq N.
\end{equation}
By reviewing the boundary conditions of Maxwell's Equations \cite{J.Xu_STAR-RIS}, the phase-shift coefficients satisfy
\begin{equation}
\label{2}
\mid\theta^{\text{t}}_{n}-\theta^{\text{r}}_{n}\mid=\frac{\pi}{2}\ \text{or}\ \frac{3\pi}{2}, \ 1\leq n\leq N.
\end{equation}
As such, for an $N$-element STAR-RIS, the transmission/reflection coefficient matrix is given by
\begin{equation}
\label{3}
\bm{\Theta}_{\text{t}/\text{r}}=\text{diag}([\sqrt{\beta^{\text{t}/\text{r}}_{1}}e^{j\theta^{\text{t}/\text{r}}_{1}},\dots,
\sqrt{\beta^{\text{t}/\text{r}}_{N}}e^{j\theta^{\text{t}/\text{r}}_{N}}]).
\end{equation}

\subsection{Signal Model}
The BS exploits the multiple beamforming vectors $\mathbf{w}_{\text{I}},\mathbf{w}_{\text{O}}\in\mathbb{C}^{M\times 1}$ to send the confidential signals $s_{\text{I}}$ and $s_{\text{O}}$ ($\mathbb{E}\{|s_{\text{I}}|^{2}\}=\mathbb{E}\{|s_{\text{O}}|^{2}\}=1$) to IU and OU, respectively. Thus, the received signals at IU and $\text{E}_{1}$ are given by
\begin{equation}
\label{4}
y_{\varsigma}=\mathbf{h}_{\varsigma,\text{S}}^{H}\bm{\Theta}_{\text{t}}^{H}\mathbf{G}\big(\sum\nolimits_{\varrho\in\{\text{I},\text{O}\}}\mathbf{w}_{\varrho}s_{\varrho}\big)
+n_{\varsigma},\ \varsigma\in\{\text{I},\text{E}_{1}\},
\end{equation}
where $n_{\text{I}},n_{\text{E}_{1}}\sim \mathcal{CN}(0,\sigma^{2})$ represent the additive white Gaussian noise (AWGN) at the IU and $\text{E}_{1}$. Similarly, the received signals at OU and $\text{E}_{2}$ are given by
\begin{equation}
\label{5}
y_{\varsigma}=\mathbf{h}_{\varsigma,\text{S}}^{H}\bm{\Theta}_{\text{r}}^{H}\mathbf{G}\big(\sum\nolimits_{\varrho\in\{\text{I},\text{O}\}}\mathbf{w}_{\varrho}s_{\varrho}\big)
+n_{\varsigma},\ \varsigma\in\{\text{O},\text{E}_{2}\}.
\end{equation}
On receiving superimposed signals, each legitimate user only decodes its own signal while treating the other's signal as interference. However, due to the full-space propagation characteristic of STAR-RIS, a new type of eavesdropping, i.e., mutual eavesdropping, exists, which indicates that both $\text{E}_{1}$ and $\text{E}_{2}$ can not only wiretap the signal intended for the user on the same side of STAR-RIS, but also can access the signal for opposite-side user. Hence, the achievable rates for legitimate users and eavesdroppers to decode signals they are interested in are given by
\begin{equation}
\label{6}
R_{\varrho}=\log_{2}\left(1+\frac{|\mathbf{h}_{\varrho,\text{S}}^{H}\bm{\Theta}_{\text{t}/\text{r}}^{H}\mathbf{G}\mathbf{w}_{\varrho}|^{2}}
{|\mathbf{h}_{\overline{\varrho},\text{S}}^{H}\bm{\Theta}_{\text{t}/\text{r}}^{H}\mathbf{G}\mathbf{w}_{\overline{\varrho}}|^{2}+\sigma^{2}}\right),
\end{equation}
\begin{equation}
\label{7}
R_{\text{E}_{k},\varrho}=\log_{2}\left(1+\frac{|\mathbf{h}_{\text{E}_{k},\text{S}}^{H}\bm{\Theta}_{\text{t}/\text{r}}^{H}\mathbf{G}\mathbf{w}_{\varrho}|^{2}}
{|\mathbf{h}_{\text{E}_{k},\text{S}}^{H}\bm{\Theta}_{\text{t}/\text{r}}^{H}\mathbf{G}\mathbf{w}_{\overline{\varrho}}|^{2}+\sigma^{2}}\right),
\end{equation}
where $k\in\{1,2\}$, $\varrho\in\{\text{I},\text{O}\}$, and $\overline{\varrho}\in\{\text{I},\text{O},\overline{\varrho}\neq\varrho\}$.

Accordingly, by adopting the wiretap code scheme, the secrecy capacity of IU or OU is given by
\begin{equation}
\label{8}
R_{\text{s},\varrho}=\left[R_{\varrho}-\max\{R_{\text{E}_{1},\varrho},R_{\text{E}_{2},\varrho}\}\right]^{+},\ \varrho\in\{\text{I},\text{O}\},
\end{equation}
where $[x]^{+}=\max\{x,0\}$.

\subsection{Problem Formulation}
In this paper, we target for maximizing the minimum secrecy capacity of legitimate users by jointly optimizing transmit beamforming at the BS and coupled transmission and reflection coefficients at the STAR-RIS, subject to the transmit power budget and amplitude/phase-shift coupling constraints. The problem is formulated as follows.
\begin{subequations}
\begin{align}
\label{9a} &\max\limits_{\mathbf{w}_{\text{I}},\mathbf{w}_{\text{O}},\bm{\Theta}^{\text{t}},\bm{\Theta}^{\text{r}}}\quad \min\limits_{\varrho\in\{\text{I},\text{O}\}}\ R_{\text{s},\varrho},\\
\label{9b}&\quad\text{s.t.} \quad \sum\nolimits_{\varrho\in\{\text{I},\text{O}\}}\|\mathbf{w}_{\varrho}\| \leq P_{\text{max}},\\
\label{9c}&\quad\quad\quad\,\, \beta^{\text{r}}_{n}+\beta^{\text{t}}_{n}= 1,\  \forall n,\\
\label{9d}&\quad\quad\quad \mid\theta^{\text{t}}_{n}-\theta^{\text{r}}_{n}\mid=\frac{\pi}{2}\ \text{or}\ \frac{3\pi}{2}, \ \theta^{\text{t}}_{n},\theta^{\text{r}}_{n}\in[0,2\pi),\ \forall n,
\end{align}
\end{subequations}
where $P_{\text{max}}$ represents the transmit power budget at the BS. Note that \eqref{9b} denotes the total power consumption constraint at the BS, while \eqref{9c} and \eqref{9d} denote the amplitude and phase-shift coupling constraints at the STAR-RIS, respectively.

\section{Proposed Solution}
\label{sec:pagestyle}
In this section, we devise a PSB algorithm to tackle the non-convex problem (9). Precisely, we first construct the AL problem with the relaxed phase-shift constraints in the outer loop, and then, the transmit beamforming, the transmission/reflection coefficients and the optimal coupled coefficients are alternately optimized by employing SDR technique and first-order optimality condition in the inner loop. Finally, an appropriate modification of the proposed algorithm is designed for the extension to the discrete coupled phase shifts.

\subsection{Outer Layer Problem Reformulation}
Here, we denote $\mathbf{u}_{\text{t}}=[\sqrt{\beta^{\text{t}}_{1}}e^{j\theta^{\text{t}}_{1}},\dots,\sqrt{\beta^{\text{t}}_{N}}e^{j\theta^{\text{t}}_{N}}]^{T}$, $\mathbf{u}_{\text{r}}=[\sqrt{\beta^{\text{r}}_{1}}e^{j\theta^{\text{r}}_{1}},\dots,\sqrt{\beta^{\text{r}}_{N}}e^{j\theta^{\text{r}}_{N}}]^{T}$,
$\mathbf{V}_{\varrho}=\mathbf{G}^{H}\text{diag}(\mathbf{h}_{\varrho,\text{S}})$ and $\mathbf{V}_{\text{E}_{k}}=\mathbf{G}^{H}\text{diag}(\mathbf{h}_{\text{E}_{k},\text{S}})$, where $\varrho\in\{\text{I},\text{O}\}$ and $k\in\{1,2\}$. Thus, the secrecy capacity can be equivalently expressed as
\begin{small}
\begin{align}
\label{10}\nonumber
R_{\text{s},\varrho}=&\Bigg[\log_{2}\left(1+\frac{\text{Tr}(\mathbf{W}_{\varrho}\mathbf{V}_{\varrho}\mathbf{U}_{\text{t}/\text{r}}\mathbf{V}_{\varrho}^{H})}
{\text{Tr}(\mathbf{W}_{\overline{\varrho}}\mathbf{V}_{\varrho}\mathbf{U}_{\text{t}/\text{r}}\mathbf{V}_{\varrho}^{H})+\sigma^{2}}\right)-\\
&\max\limits_{k\in\{1,2\}}\log_{2}\left(1+\frac{\text{Tr}(\mathbf{W}_{\varrho}\mathbf{V}_{\text{E}_{k}}\mathbf{U}_{\text{t}/\text{r}}\mathbf{V}_{\text{E}_{k}}^{H})}
{\text{Tr}(\mathbf{W}_{\overline{\varrho}}\mathbf{V}_{\text{E}_{k}}\mathbf{U}_{\text{t}/\text{r}}\mathbf{V}_{\text{E}_{k}}^{H})+\sigma^{2}}\right)\Bigg]^{+},
\end{align}\end{small}
\!\!\!where the semi-positive matrices satisfy $\mathbf{W}_{\varrho}=\mathbf{w}_{\varrho}\mathbf{w}_{\varrho}^{H}$ and $\mathbf{U}_{\text{t}/\text{r}}=\mathbf{u}_{\text{t}/\text{r}}\mathbf{u}_{\text{t}/\text{r}}^{H}$. To tackle the non-convexity of \eqref{10}, a linear lower bound expression of \eqref{10} is construct by introducing slack variables $\{l_{\text{n},\varrho},l_{\text{d},\varrho},e_{\text{n},\text{E}_{k},\varrho},e_{\text{d},\text{E}_{k},\varrho},\mu_{n,\varrho},
\nu_{\text{E}_{k},\varrho}\}$. Precisely, let
\begin{equation}
\label{11}
2^{l_{\text{n},\varrho}}\leq\text{Tr}(\mathbf{W}_{\varrho}\mathbf{V}_{\varrho}\mathbf{U}_{\text{t}/\text{r}}\mathbf{V}_{\varrho}^{H})+
\text{Tr}(\mathbf{W}_{\overline{\varrho}}\mathbf{V}_{\varrho}\mathbf{U}_{\text{t}/\text{r}}\mathbf{V}_{\varrho}^{H})+\sigma^{2},
\end{equation}
\begin{equation}
\label{12}
2^{l_{\text{d},\varrho}}\geq \mu_{n,\varrho},
\end{equation}
\begin{equation}
\label{13}
\mu_{\text{n},\varrho}\geq\text{Tr}(\mathbf{W}_{\overline{\varrho}}\mathbf{V}_{\varrho}\mathbf{U}_{\text{t}/\text{r}}\mathbf{V}_{\varrho}^{H})+\sigma^{2}.
\end{equation}
Thus, $\underline{R}_{\varrho}= l_{\text{n},\varrho}-l_{\text{d},\varrho}\leq R_{\varrho}$. Similarly, we define $\overline{R}_{\text{E}_{k},\varrho}=e_{\text{n},\text{E}_{k},\varrho}-e_{\text{d},\text{E}_{k},\varrho}\geq R_{\text{E}_{k},\varrho}$, where $e_{\text{n},\text{E}_{k},\varrho}$ and $e_{\text{d},\text{E}_{k},\varrho}$ satsify
\begin{equation}
\label{14}
2^{e_{\text{d},\text{E}_{k},\varrho}}\leq\text{Tr}(\mathbf{W}_{\overline{\varrho}}\mathbf{V}_{\text{E}_{k}}\mathbf{U}_{\text{t}/\text{r}}\mathbf{V}_{\text{E}_{k}}^{H})+\sigma^{2},
\end{equation}
\begin{equation}
\label{15}
\nu_{\text{E}_{k},\varrho}\!\geq\! \text{Tr}(\mathbf{W}_{\varrho}\mathbf{V}_{\text{E}_{k}}\mathbf{U}_{\text{t}/\text{r}}\mathbf{V}_{\text{E}_{k}}^{H})\!+\!
\text{Tr}(\mathbf{W}_{\overline{\varrho}}\mathbf{V}_{\text{E}_{k}}\mathbf{U}_{\text{t}/\text{r}}\mathbf{V}_{\text{E}_{k}}^{H})\!+\!\sigma^{2},
\end{equation}
\begin{equation}
\label{16}
2^{e_{\text{n},\text{E}_{k},\varrho}}\geq \nu_{\text{E}_{k},\varrho}.
\end{equation}
As a result, we can rewrite the problem (9) as
\begin{subequations}
\begin{align}
\label{17a} &\max\limits_{\mathbf{W}_{\text{I}},\mathbf{W}_{\text{O}},\mathbf{U}_{\text{t}},\mathbf{U}_{\text{r}},\mu_{n,\varrho},\nu_{\text{E}_{k},\varrho},
\atop
l_{\text{n},\varrho},l_{\text{d},\varrho},e_{\text{n},\text{E}_{k},\varrho},e_{\text{d},\text{E}_{k},\varrho}}\  \min\limits_{\varrho\in\{\text{I},\text{O}\}}\ \underline{R}_{\varrho}-\overline{R}_{\text{E}_{\text{max}},\varrho},\\
\label{17b}&\quad\text{s.t.} \quad \sum\nolimits_{\varrho\in\{\text{I},\text{O}\}}\text{Tr}(\mathbf{W}_{\varrho}) \leq P_{\text{max}},\\
\label{17c}&\quad\quad\quad\,\, \mathbf{U}_{\text{t}}(n,n)+\mathbf{U}_{\text{r}}(n,n) = 1,\ \forall n,\\
\label{17d}&\quad\quad\quad\,\, \overline{R}_{\text{E}_{k},\varrho} \leq \overline{R}_{\text{E}_{\text{max}},\varrho},\ \ \underline{R}_{\varrho}\geq \overline{R}_{\text{E}_{\text{max}},\varrho},\ \forall k,\varrho,\\
\label{17e}&\quad\quad\quad\,\, \mathbf{W}_{\varrho}\succeq\mathbf{0},\ \text{rank}(\mathbf{W}_{\varrho})=1,\ \forall \varrho,\\
\label{17f}&\quad\quad\quad\,\,  \mathbf{U}_{\text{t}/\text{r}}\succeq\mathbf{0}, \ \text{rank}(\mathbf{U}_{\text{t}/\text{r}})=1,\\
\label{17g}&\quad\quad\quad\,\, \eqref{9d},\eqref{11}-\eqref{16}.
\end{align}
\end{subequations}
To tackle the non-convex constraints \eqref{12} and \eqref{16}, we construct a linear upper bound approximation function $\mathcal{H}(x,y)$ for any logarithmic function $\log_{2}(x)$. Particularly, we define $ \mathcal{H}(x,y) \triangleq \log_{2}(y)+\frac{x-y}{y\log(2)} \geq \log_{2}(x)$, where $\mathcal{H}(x,y)$ reaches the minimum point only when $y=x$ satisfies. Thus, the non-convex constraints \eqref{12} and \eqref{16} can be converted to
\begin{equation}
\label{18}
l_{\text{d},\varrho} \geq \mathcal{H}(\mu_{n,\varrho},y_{\varrho}), \quad e_{\text{n},\text{E}_{k},\varrho} \geq \mathcal{H}(\nu_{\text{E}_{k},\varrho},y_{k,\varrho}).
\end{equation}
Furthermore, for the non-convex constraints \eqref{9d}, we consider exploiting the penalty based approach to relax the phase coupling to the auxiliary variables $\tilde{\mathbf{u}}_{\text{t}/\text{r}}=[\sqrt{\tilde{\beta}^{\text{t}}_{1}}e^{j\tilde{\theta}^{\text{t}/\text{r}}_{1}},\dots,
\sqrt{\tilde{\beta}^{\text{t}/\text{r}}_{N}}e^{j\tilde{\theta}^{\text{t}/\text{r}}_{N}}]^{T}$. Specifically, by transforming the equality constraint $\tilde{\mathbf{u}}_{\text{t}/\text{r}}=\mathbf{u}_{\text{t}/\text{r}}$ as the penalty term in the objective function \eqref{17a}, we can reformulate the original problem (9) as the following AL problem of problem (9)
\begin{subequations}
\begin{align}
\label{19a} &\max\limits_{\mathbf{W}_{\text{I}},\mathbf{W}_{\text{O}},\mathbf{U}_{\text{t}},\mathbf{U}_{\text{r}},\mu_{n,\varrho},
\atop
\nu_{\text{E}_{k},\varrho},l_{\text{n},\varrho},l_{\text{d},\varrho},e_{\text{n},\text{E}_{k},\varrho},e_{\text{d},\text{E}_{k},\varrho}}\  \min\limits_{\varrho\in\{\text{I},\text{O}\}}\ \mathfrak{F}_{\varrho},\\
\label{19b}&\quad\text{s.t.} \quad \tilde{\mathbf{u}}_{\text{r}}(n) = \pm j\tilde{\mathbf{u}}_{\text{t}}(n),\quad \sum\nolimits_{\varsigma\in\{\text{t},\text{r}\}}|\tilde{\mathbf{u}}_{\varsigma}(n)|^{2}= 1,\\
\label{19c}&\quad\quad\quad\,\, \eqref{11},\eqref{13}-\eqref{15},\eqref{17b}-\eqref{17f},\eqref{18},
\end{align}
\end{subequations}
where $\mathfrak{F}_{\varrho} = \underline{R}_{\varrho}-\overline{R}_{\text{E}_{\text{max}},\varrho}-\frac{1}{2\rho}(\sum_{\varsigma\in\{\text{t},\text{r}\}}\|\tilde{
\mathbf{u}}_{\varsigma}\tilde{\mathbf{u}}_{\varsigma}^{H}-\mathbf{U}_{\varsigma}+\rho\bm{\lambda}_{\varsigma}\|^{2})$. Note that $\bm{\lambda}_{\varsigma}$ denotes the lagrangian dual variable, while $\rho>0$ is the penalty scaling factor.

\subsection{Inner Layer Problem Optimization}

\textit{1) Transmit beamforming optimization:}  For any given $\{\mathbf{U}_{\text{t}/\text{r}},\tilde{\mathbf{u}}_{\text{t}/\text{r}}\}$, the active beamforming $\{\mathbf{W}_{\text{I}},\mathbf{W}_{\text{O}}\}$ can be optimized via solving following problem
\begin{subequations}
\begin{align}
\label{20a} &\max\limits_{\mathbf{W}_{\text{I}},\mathbf{W}_{\text{O}},\mu_{n,\varrho},\nu_{\text{E}_{k},\varrho},l_{\text{n},\varrho},
\atop
l_{\text{d},\varrho},e_{\text{n},\text{E}_{k},\varrho},e_{\text{d},\text{E}_{k},\varrho}}\  \min\limits_{\varrho\in\{\text{I},\text{O}\}}\ \underline{R}_{\varrho}-\overline{R}_{\text{E}_{\text{max}},\varrho},,\\
\label{20b}&\quad\text{s.t.} \quad \eqref{11},\eqref{13}-\eqref{15},\eqref{17b},\eqref{17d},\eqref{17e}.
\end{align}
\end{subequations}
Problem (20) can be efficiently solved by iteratively updating $y_{\varrho}=\mu_{n,\varrho}$ and $y_{k,\varrho}=\nu_{\text{E}_{k},\varrho}$. Note that the rank-one constraints can be reasonably dropped because the rank of $\mathbf{W}_{\varrho}$ only relies on that of $\mathbf{U}_{\text{t}/\text{r}}$, i.e., $\text{rank}(\mathbf{W}_{\varrho})\leq\text{rank}(\mathbf{U}_{\text{t}/\text{r}})$=1, which is proved in Appendix A.

\textit{2) Relaxed transmission/reflection coefficient optimization}: With the fixed $\{\mathbf{W}_{\text{I}},\mathbf{W}_{\text{O}}\}$, the transmission and reflection coefficients $\{\mathbf{U}_{\text{t}/\text{r}}\}$ can be optimized through solving following problem
\begin{subequations}
\begin{align}
\label{21a} &\max\limits_{\mathbf{U}_{\text{t}},\mathbf{U}_{\text{r}},\mu_{n,\varrho},\nu_{\text{E}_{k},\varrho},
\atop
l_{\text{n},\varrho},l_{\text{d},\varrho},e_{\text{n},\text{E}_{k},\varrho},e_{\text{d},\text{E}_{k},\varrho}}\  \min\limits_{\varrho\in\{\text{I},\text{O}\}}\ \mathfrak{F}_{\varrho},\\
\label{21b}&\quad\text{s.t.} \quad \eqref{11},\eqref{13}-\eqref{15},\eqref{17c},\eqref{17f}.
\end{align}
\end{subequations}
Here, we adopt the penalty based difference-of-convex (DC) method for extracting the $\mathbf{u}_{\text{t}}$ and $\mathbf{u}_{\text{r}}$. Precisely, by equivalently converting $\text{rank}(\mathbf{U}_{\text{t}/\text{r}})=1$ to $\text{Tr}(\mathbf{U}_{\text{t}/\text{r}})=\|\mathbf{U}_{\text{t}/\text{r}}\|_{2}$, the objective function can be replaced by following DC form with dropping the rank-one constraints \cite{T.Jiang_DCP}.
\begin{equation}
\label{22}
\mathfrak{F}_{\varrho}^{\text{p}}=\mathfrak{F}_{\varrho}-\tau\Big(\sum\nolimits_{\varsigma\in\{\text{t},\text{r}\}}
\Re\Big(\text{Tr}(\mathbf{U}_{\varsigma}^{H}(\mathbf{I}-\mathbf{u}_{\varsigma,1}\mathbf{u}_{\varsigma,1}^{H}))\Big)\Big),
\end{equation}
where $\Re(\cdot)$ denotes the real part of the complex number, and $\tau>0$ denotes the scaling coefficient of rank-one penalty terms. As results, transmission/reflection coefficients can be directly obtained in an iterative manner \cite{ZZ_STAR}.

\textit{3) Coupled transmission/reflection coefficient optimization:} With the fixed rank-one matrices $\{\mathbf{W}_{\text{I}},\mathbf{W}_{\text{O}},\mathbf{U}_{\text{t}/\text{r}}\}$, the auxiliary variables $\{\tilde{\mathbf{u}}_{\text{t}/\text{r}}\}$ can be optimized by solving following equivalent problem
\begin{subequations}
\begin{align}
\label{23a} &\min\limits_{\tilde{\mathbf{u}}_{\varsigma}}\  \sum_{\varsigma\in\{\text{t},\text{r}\}}
\|\tilde{\mathbf{u}}_{\varsigma}-\mathbf{u}_{\varsigma}\|^{2},\\
\label{23b}&\text{s.t.} \ \  \eqref{19b}.
\end{align}
\end{subequations}
Since optimization variables $\{\tilde{\mathbf{u}}_{\varsigma}(n),\tilde{\mathbf{u}}_{\varsigma}(m)\}$ ($n\neq m$) are absolutely separable in the objective function, the problem (23) can be optimized by solving $N$ independent subproblems with respect to $\{\tilde{\mathbf{u}}_{\text{t}}(n),\tilde{\mathbf{u}}_{\text{r}}(n)\}$. With some  algebraic manipulations, the resultant subproblem is expressed as
\begin{subequations}
\begin{align}
\label{24a} &\max\limits_{\tilde{\mathbf{u}}_{\varsigma}(n)}\  \Re(\mathbf{u}_{\text{t}}^{H}(n)\tilde{\mathbf{u}}_{\text{t}}(n))+\Re(\mathbf{u}_{\text{r}}^{H}(n)\tilde{\mathbf{u}}_{\text{r}}(n)),\\
\label{24b}&\quad\text{s.t.} \quad \eqref{19b}.
\end{align}
\end{subequations}
Then, with any fixed $\tilde{\beta}^{\text{t}/\text{r}}_{n}$, the objective function \eqref{24a} can be rewritten as $\Re([\mathbf{v}_{\text{t}}^{H}(n)\pm j \mathbf{v}_{\text{t}}^{H}(n)]e^{j\tilde{\theta}^{\text{t}}_{n}})$, where $\mathbf{v}_{\varsigma}^{H}=\mathbf{u}_{\varsigma}^{H}\text{diag}([\sqrt{\tilde{\beta}^{\varsigma}_{1}},\cdots,\sqrt{\tilde{\beta}^{\varsigma}_{N}}])$.
It reaches maximum only when $\angle([\mathbf{v}_{\text{t}}^{H}(n)\pm j \mathbf{v}_{\text{t}}^{H}(n)]e^{j\tilde{\theta}^{\text{t}}_{n}})=0$ holds, yielding the following optimal solution
\begin{align}
\label{25}
\!\!\!\tilde{\theta}^{\text{t},\text{opt}}_{n}\!=\!
\begin{cases}
        \!-\angle(\mathbf{v}_{\text{t}}^{H}(n)\!+\! j \mathbf{v}_{\text{t}}^{H}(n)),\ \! (\tilde{\theta}^{\text{r},\text{opt}}_{n}\!\!=\!\tilde{\theta}^{\text{t},\text{opt}}_{n}\!+\!\frac{\pi}{2}),\\
        \!-\angle(\mathbf{v}_{\text{t}}^{H}(n)\!-\! j \mathbf{v}_{\text{t}}^{H}(n)),\ \! (\tilde{\theta}^{\text{r},\text{opt}}_{n}\!\!=\!\tilde{\theta}^{\text{t},\text{opt}}_{n}\!+\!\frac{3\pi}{2}).
        \end{cases}
\end{align}
Similarly, with any fixed $\tilde{\theta}^{\text{t}/\text{r}}_{n}$, objective function \eqref{24a} is equivalent to $\Re(\bm{\psi}_{\text{t}}^{H}(n)\sqrt{\tilde{\beta}^{\text{t}}_{n}}+\bm{\psi}_{\text{r}}^{H}(n)\sqrt{\tilde{\beta}^{\text{r}}_{n}})$, where $\bm{\psi}_{\varsigma}^{H}=\\ \mathbf{u}_{\varsigma}^{H}\text{diag}([e^{j\tilde{\theta}^{\varsigma}_{n}},\cdots,e^{j\tilde{\theta}^{\varsigma}_{N}}])$. Let $p_{n}=|\bm{\psi}_{\text{t}}^{H}(n)|\cos(\angle\bm{\psi}_{\text{t}}^{H}(n))$ and $q_{n}=|\bm{\psi}_{\text{r}}^{H}(n)|\cos(\angle\bm{\psi}_{\text{r}}^{H}(n))$, the objective function can be further expressed as $p_{n}\sqrt{\tilde{\beta}^{\text{t}}_{n}}+q_{n}\sqrt{\tilde{\beta}^{\text{r}}_{n}}$. Then, by checking the first-order optimality condition, the optimal amplitude coefficients are given by
\begin{align}
\label{26}
\!\!\!\!\!\begin{cases}
        \sqrt{\tilde{\beta}^{\text{t},\text{opt}}_{n}}\!=\!\frac{p_{n}}{\sqrt{p^{2}_{n}+q^{2}_{n}}},
        \sqrt{\tilde{\beta}^{\text{r},\text{opt}}_{n}}\!=\!\frac{q_{n}}{\sqrt{p^{2}_{n}+q^{2}_{n}}}, \text{if}\  p_{n},\!q_{n}\geq 0;\\
        \sqrt{\tilde{\beta}^{\text{t},\text{opt}}_{n}}=1,\sqrt{\tilde{\beta}^{\text{r},\text{opt}}_{n}}=0, \ \text{if}\  p_{n}\geq 0, q_{n}< 0;\\
        \sqrt{\tilde{\beta}^{\text{t},\text{opt}}_{n}}=0,\sqrt{\tilde{\beta}^{\text{r},\text{opt}}_{n}}=1, \ \text{if}\  p_{n}< 0, q_{n}\geq 0;\\
        \sqrt{\tilde{\beta}^{\text{t},\text{opt}}_{n}}=\sqrt{\tilde{\beta}^{\text{r},\text{opt}}_{n}}=0,\  \text{otherwise}.
        \end{cases}
\end{align}

\begin{table}[t]
    \centering
    \begin{tabular}{p{225pt}}
    \toprule
    \textbf{Algorithm-1:} PSB Algorithm \\
    \midrule
    1: Initialize the iteration parameters with $l=1$;\\
    2:  \textbf{Outer layer repeat}\\
    3: \quad\textbf{Inner layer repeat }\\
    4:         \quad \quad Optimize $\{\mathbf{W}_{\text{I}}^{l},\mathbf{W}_{\text{O}}^{l}\}$ by solving problem (20);\\
    5:         \quad \quad Optimize $\{\mathbf{U}_{\text{t}}^{l},\mathbf{U}_{\text{r}}^{l}\}$ by solving problem (21);\\
    6:         \quad \quad Update $\{\tilde{\mathbf{u}}_{\text{t}}^{l},\tilde{\mathbf{u}}_{\text{r}}^{l}\}$ by \eqref{25} and \eqref{26};\\
    7:  \quad\textbf{If} $\mathfrak{V}(\tilde{\mathbf{u}}_{\varsigma}^{l},\mathbf{U}_{\varsigma}^{l}) \leq c_{1}\mathfrak{V}(\tilde{\mathbf{u}}_{\varsigma}^{l-1},\mathbf{U}_{\varsigma}^{l-1})$  with $c_{1}<1$;\\
    8: \qquad $\bm{\lambda}_{\varsigma}^{l+1}=\bm{\lambda}_{\varsigma}^{l}+\frac{1}{\rho}[\tilde{\mathbf{U}}_{\varsigma}^{l+1}-\mathbf{U}_{\varsigma}^{l+1}]$;\\
    9:  \quad\textbf{Else} $\rho^{l} = c_{2}\rho^{l-1}$ with $c_{2}<1$;\\
    10: \ \  Set $l=l+1$, ;\\
    11: \textbf{Until} converge with the accuracy $\varepsilon_{\text{th}}$;\\
    \bottomrule
    \end{tabular}
\end{table}

\subsection{Overall Algorithm}
The overall algorithm is summarized in \textbf{Algorithm-1}, where $\mathfrak{V}(\tilde{\mathbf{u}}_{\varsigma}^{l},\mathbf{U}_{\varsigma}^{l})\triangleq
\|\tilde{\mathbf{u}}_{\varsigma}^{l}(\tilde{\mathbf{u}}_{\varsigma}^{H})^{l}-\mathbf{U}_{\varsigma}^{l}\|_{\infty}$, and $\varepsilon_{\text{th}}$ denotes the convergence accuracy. Since the optimal solutions $\tilde{\mathbf{u}}_{\text{t}/\text{r}}$ and stationary point solutions $\{\mathbf{W}_{\text{I}},\mathbf{W}_{\text{O}}\}$ and $\{\mathbf{U}_{\text{t}/\text{r}}\}$ are obtained at each step, the objective function is ensured to be non-decreasing over the inner loop iterations. Then, by updating the Lagrangian dual variable $\bm{\lambda}_{\varsigma}$ or decreasing penalty scaling factor $\rho$ in the out loop, the proposed PSB algorithm would converge the stationary point solutions of the original problem \cite{PDD-1}.

The overall computational complexity of \textbf{Algorithm-1} is mainly determined by solving the SDP with respect to $\{\mathbf{W}_{\text{I}},\mathbf{W}_{\text{O}}\}$ and $\{\mathbf{U}_{\text{t}/\text{r}}\}$ in the inner loop. By exploiting the interior point method, the overall complexity is given by $\mathcal{O}\left(\log(\frac{1}{\varepsilon_{\text{th}}})(l_{\text{a}}2M^{3.5}+l_{\text{p}}2N^{3.5})\right)$, where $\mathcal{O}$ denotes the big-$\mathcal{O}$ notation, and $l_{\text{a}}$ and $l_{\text{p}}$ denote the iteration numbers for solving problem (20) and (21).

\subsection{Extension to Discrete Coupled Phase Shifts}
Generally, discrete phase-shift adjustment is practical and realistic for STAR-RIS, which yields a uniform quantized phase-shift feasible region, i.e.,
\begin{equation}
\label{27}
\theta^{\text{t},\text{d}}_{n},\theta^{\text{r},\text{d}}_{n}\in\mathfrak{Q}=\left\{0,\frac{2\pi}{2^{q}},\dots,\frac{2\pi(2^{q}-1)}{2^{q}}\right\}, \  \forall n,
\end{equation}
where $q$ denotes the number of quantization bits. Note that the optimization problem in discrete phase-shift case possesses the same structure as the continuous phase-shift case except for discrete phase-shift constraint. Thus, we can first perform the PSB algorithm to obtain the solution $\theta^{\text{t}/\text{r},\text{c}}_{n}$ with continuous phase shifters. Then, the optimal discrete phase shifts can be obtained by $\theta^{\text{t}/\text{r},\text{d},\text{opt}}_{n}=\text{arg} \min\limits_{\theta^{\text{t}/\text{r},\text{d}}_{n}\in\mathfrak{Q}}\mid\theta^{\text{t}/\text{r},\text{d}}_{n}-\theta^{\text{t}/\text{r},\text{c}}_{n}\mid$.

\section{Numerical Results}\label{4:Num-s}
\begin{figure}[t]
\centering
\begin{minipage}[b]{0.24\textwidth} 
\centering 
\includegraphics[width=1\textwidth]{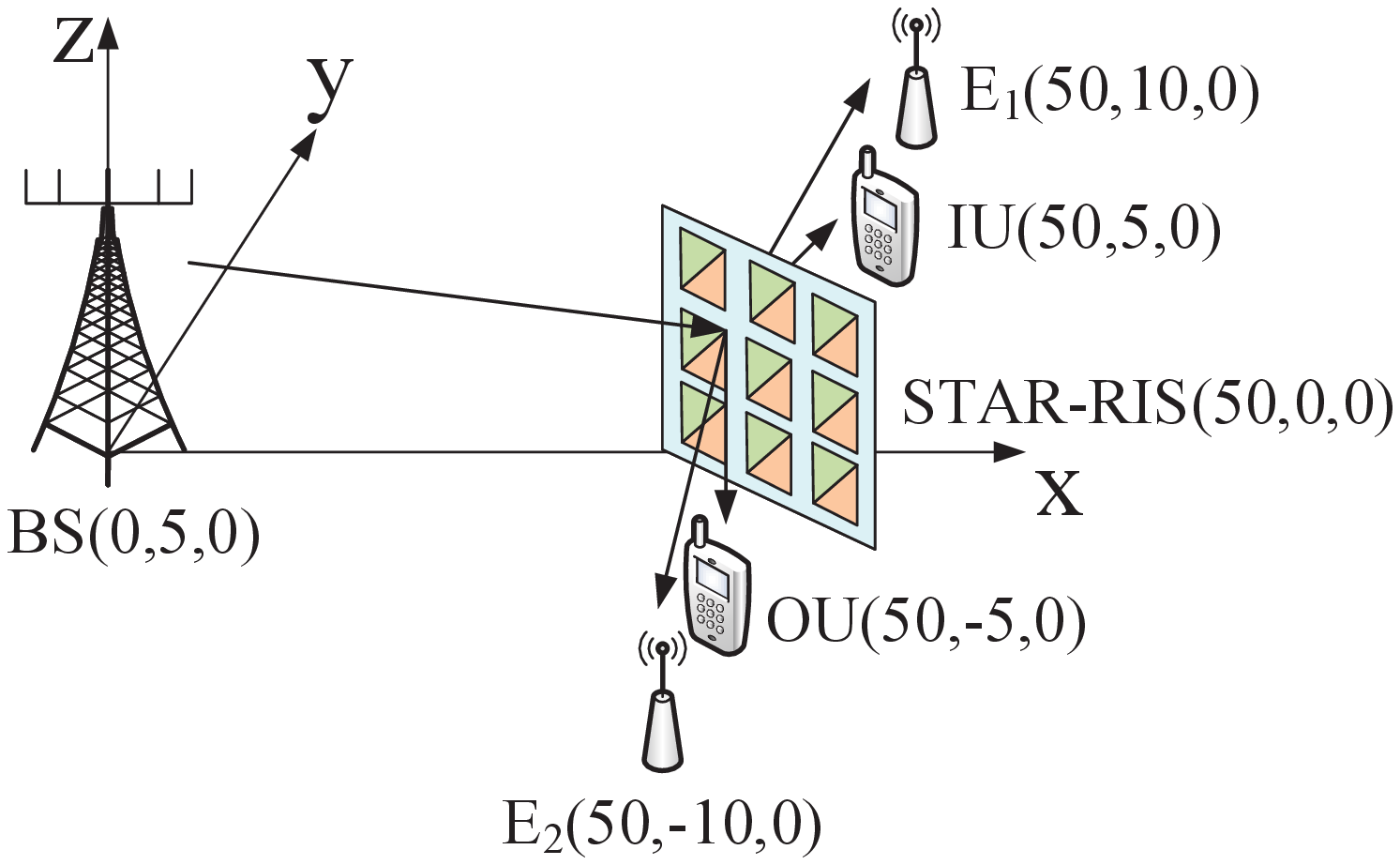} 
\caption{Simulation setup for the considered network.}
\label{fig2:sim-model}
\end{minipage}
\begin{minipage}[b]{0.24\textwidth} 
\centering 
\includegraphics[width=1\textwidth]{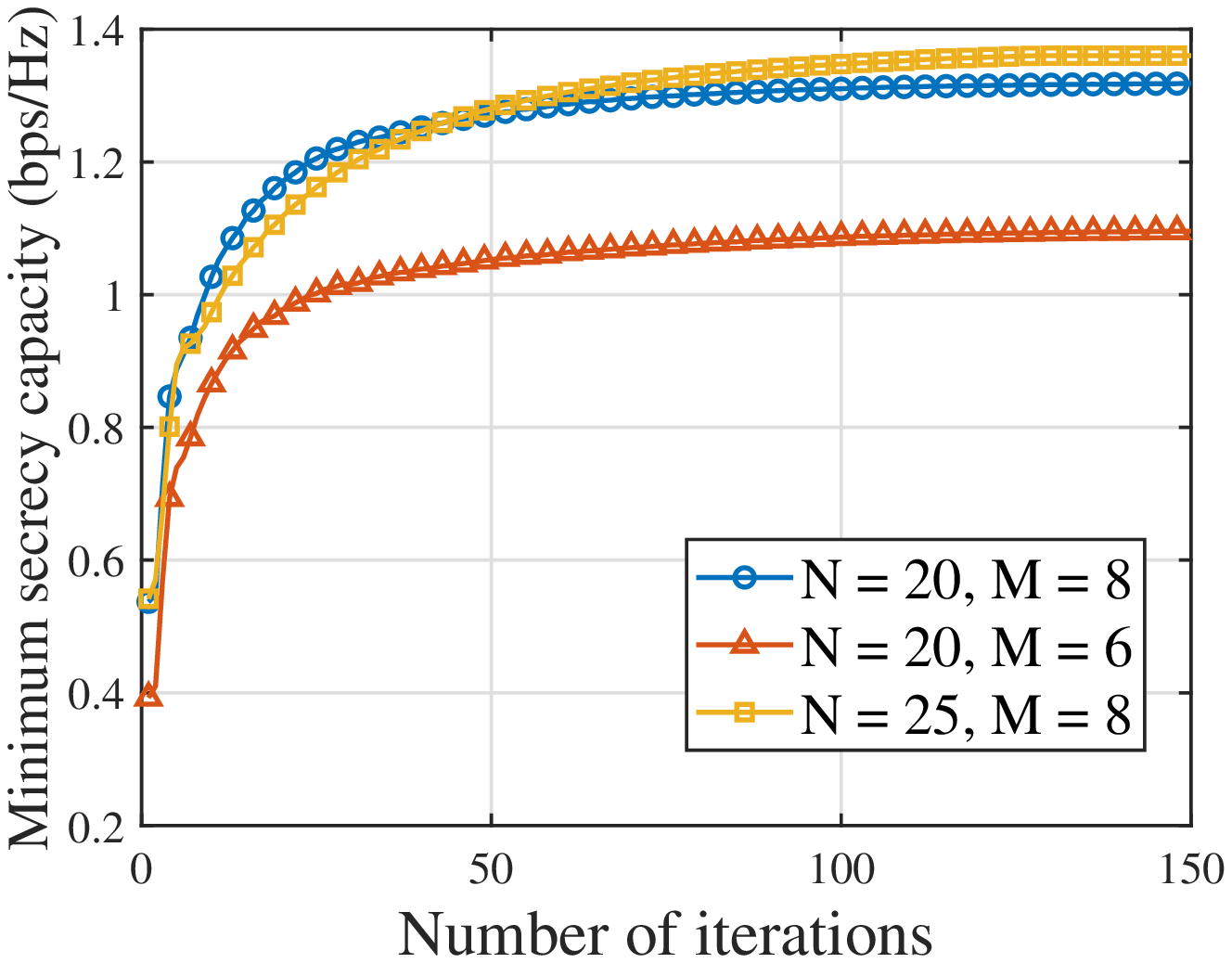}
\caption{Convergence of \textbf{Algorithm-1} with $P_{\text{max}}=-5$ dBm.}
\label{fig3:Convergence}
\end{minipage}
\end{figure}

In this section, numerical results are provided to validate the performance of the proposed algorithm. As shown in Fig. \ref{fig2:sim-model}, we consider a three-dimensional coordinate system setup, where the BS is located at $(0,0,0)$ meter (m), STAR-RIS is deployed at $(50,0,0)$ m, IU and OU are located at $(50,5,0)$ m and $(50,-5,0)$ m, respectively, while $\text{E}_{1}$ and $\text{E}_{2}$ are situated at $(50,10,0)$ m and $(50,-10,0)$ m. Both the large-scale path loss and small-scale fading are considered, so the channel model is given by $\mathbf{h}=\sqrt{L_{0}d^{-{\alpha}}}(\sqrt{\frac{\kappa}{1+\kappa}}\bar{\mathbf{h}}+\sqrt{\frac{1}{1+\kappa}}\tilde{\mathbf{h}})$. Thereinto, $\bar{\mathbf{h}}$ and $\tilde{\mathbf{h}}$ respectively denote the LoS and non-LoS components of $\mathbf{h}$, $L_{0}$ represents the path loss at the unit reference distance, $\alpha$ and $d$ are the corresponding path loss exponent and the transmit distance, and $\kappa$ is the Rician factor. The main simulation parameters are set as $L_{0}=-30$ dB, $\alpha_{\text{B},\text{S}}=2.2$, $\alpha_{\text{I},\text{S}}=\alpha_{\text{O},\text{S}}=\alpha_{\text{E}_{1},\text{S}}=\alpha_{\text{E}_{2},\text{S}}=2.5$, $\sigma^{2}=-105$ dBm, $\varepsilon_{\text{th}}=10^{-3}$, $\kappa=5$ and $c_{1}=c_{2}=0.99$. Moreover, each result is averaged over $100$ independent Monte-Carlo trials.


Fig. \ref{fig3:Convergence} depicts the convergence performance of the proposed algorithm with different $M$ and $N$. We observe that the minimum secrecy capacity monotonically increases with the number of iterations and converges to stable solutions within the finite iterations. We also observe that increasing both $M$ and $N$ is conducive to improving the minimum secrecy capacity of legitimate users but an increase in $N$ slightly deteriorates the convergence performance. An intuitive explanation for this result is that the larger $M$ and $N$ would provide more spatial degrees-of-freedom (DoFs), which enables a more flexible beamforming design for secrecy enhancement. However, increasing $N$ extremely increases the algorithm complexity when $N\gg M$ holds, so more iterations are required to achieve convergence.


\begin{figure}[t]
\centering
\begin{minipage}[b]{0.24\textwidth}
\centering
\includegraphics[width=1\textwidth]{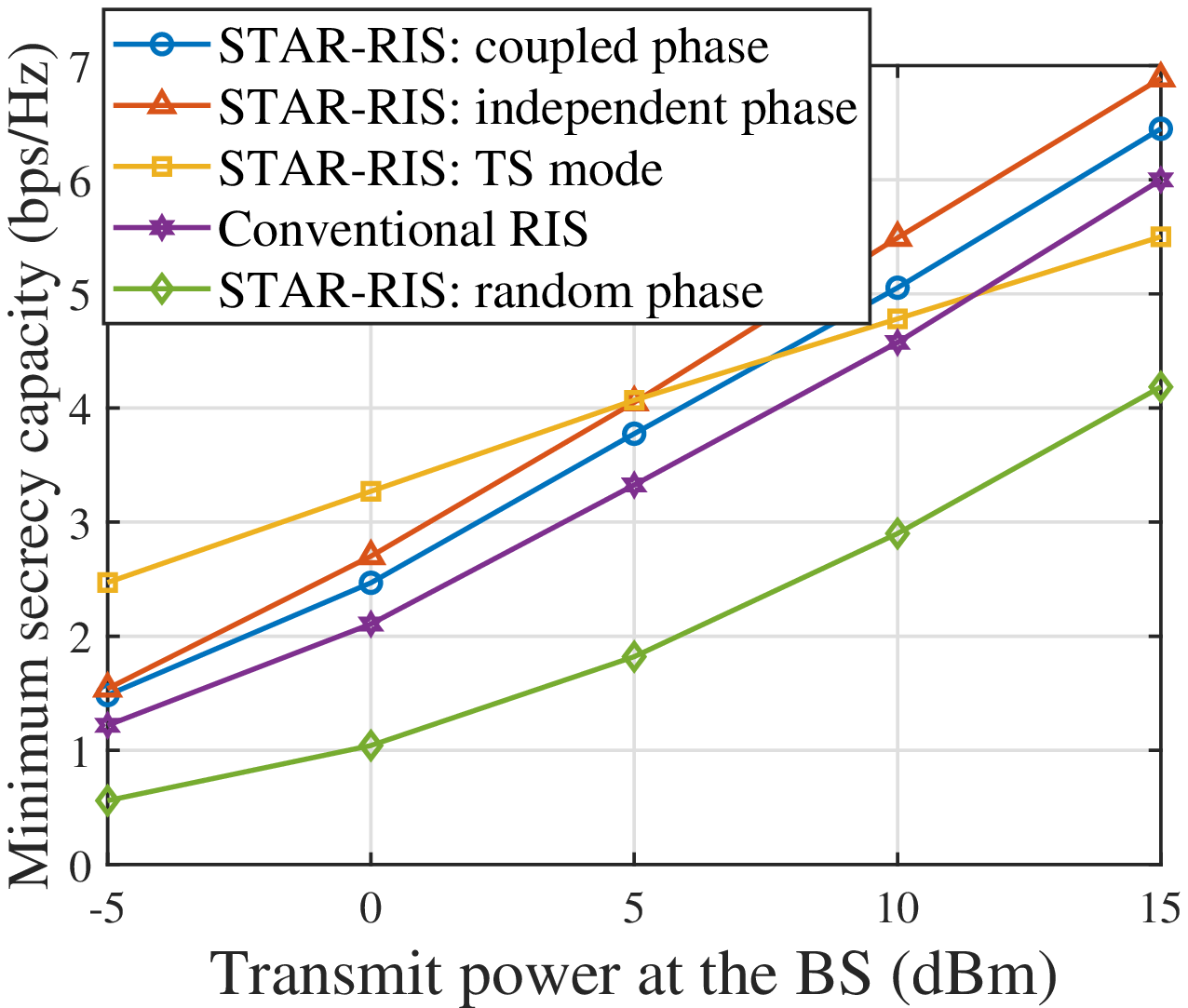}
\caption{The minimum secrecy capacity versus the transmit power budget with $M=8$ and $N=20$.}
\label{fig4:Benchmarks}
\end{minipage}
\begin{minipage}[b]{0.24\textwidth}
\centering
\includegraphics[width=1\textwidth]{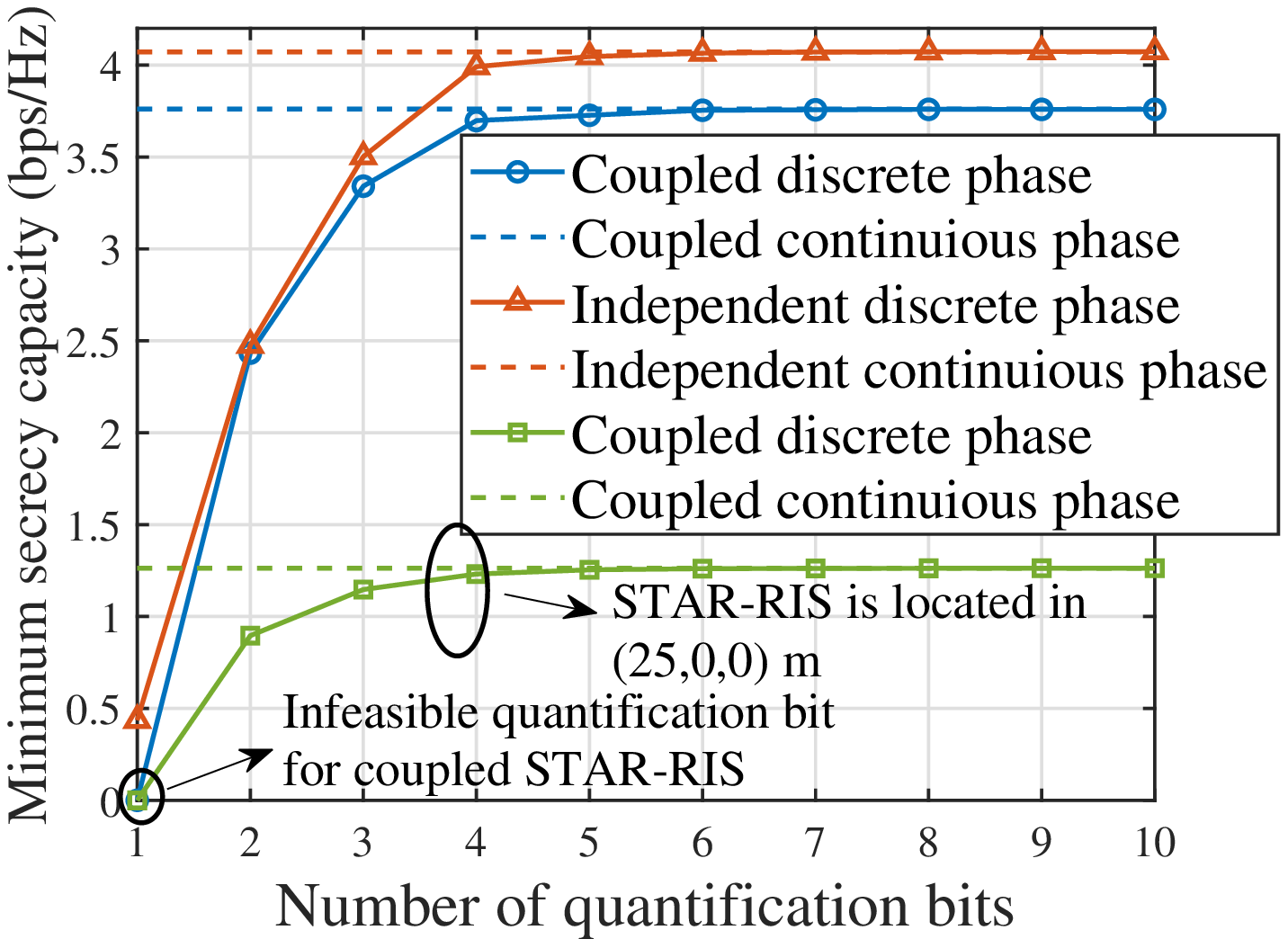}
\caption{The minimum secrecy capacity versus the number of quantification bits with $M=8$ and $N=20$.}
\label{fig5:Quantification}
\end{minipage}
\end{figure}

Fig. \ref{fig4:Benchmarks} compares the minimum secrecy capacity achieved by different schemes versus the transmit power budget, i.e., independent phase-shift STAR-RIS, time switch (TS) mode STAR-RIS, conventional RIS (C-RIS), and random phase-shift STAR-RIS. Particularly, in the TS mode scheme, STAR-RIS firstly operates in the transmission mode for supporting IU, and then switches to the reflection mode to serve OU. While in the C-RIS scheme, a reflecting-only RIS and a transmitting-only RIS are deployed at the same location as STAR-RIS with $\frac{N}{2}$ elements \cite{X.Mu_STAR-RIS}. As shown in Fig. \ref{fig4:Benchmarks}, coupled phase-shift STAR-RIS achieves a higher secrecy performance than C-RIS and the random phase shifts, which is expected since STAR-RIS exploits double DoFs than C-RIS to improve legitimate reception while minimizing information leakage, while the random phase shifts cannot always guarantee this. We also observe that due to the phase coupling, the secrecy performance achieved by coupled phase-shift STAR-RIS is slightly inferior to that of the independent phase-shift STAR-RIS. Moreover, since the TS scheme naturally avoids mutual wiretap and enjoys interference-free decoding, it achieves better security than ES STAR-RIS in the low transmit power region. But with the increasing transmit power, the time resource utilization efficiency of ES STAR-RIS becomes the dominant factor, thus achieving higher secrecy capacity in the high transmit power region.

Fig. \ref{fig5:Quantification} plots minimum secrecy capacity versus the number of the phase-shift quantization bits. We observe from Fig. \ref{fig5:Quantification} that regardless of whether the phase shifts are coupled or not, at least 4-bit quantization is required to realize nearly the same performance as continuous phase shifts. It is because different from the conventional reflecting/transmitting-only RIS, STAR-RIS needs to simultaneously strengthen the legitimate links and reduce the confidential information leakage to the eavesdroppers in an omnidirectional space, so a more exact phase-shift control is required. It is also shown that due to the existence of  ``double path loss'' for SATR-RIS aided cascaded links, deploying STAR-RIS at the user end can achieve higher secrecy performance. Moreover, since the difference between the transmission phase shift and the reflection phase shift is fixed to $\frac{\pi}{2}$/$\frac{3\pi}{2}$, so the coupled STAR-RIS is invalidated in the case of 1-bit quantification.

\section{Conclusion}\label{5:Con}
A novel secrecy beamforming scheme was proposed for coupled phase-shift STAR-RIS aided downlink communications. A PSB algorithm was developed to maximize the minimum secrecy capacity among legitimate users via jointly optimizing transmit beamforming and transmission/reflection coefficients with coupled phase shifts. Simulation results were provided to demonstrate the secrecy advantage of the proposed secrecy beamforming scheme compared to the benchmark schemes. It is also revealed that at least 4-bit quantification should be guaranteed for the coupled phase-shift STAR-RIS to achieve the comparable secrecy performance of the continuous phase-shift case.

\section*{Appendix A: Proof of the Rank-One Property}
The rank-one relaxed problem (22) is jointly convex for optimization variables and satisfies the Slater's constraint qualification. Hence, the lagrangian function with respect to $\mathbf{W}_{\varrho}$ is given by
\begin{align}
\nonumber
&\mathcal{L} = \sum\nolimits_{\varrho\in\{\text{I},\text{O}\}}\!\!\text{Tr}(\mathbf{W}_{\varrho})+\lambda_{1}
\text{Tr}(\mathbf{W}_{\varrho}\mathbf{V}_{\varrho}\mathbf{U}_{\text{t}/\text{r}}\mathbf{V}_{\varrho}^{H})-\\ \tag{A-1}
&\lambda_{2}\text{Tr}(\mathbf{W}_{\varrho}\mathbf{V}_{\text{E}_{k}}\mathbf{U}_{\text{t}/\text{r}}\mathbf{V}_{\text{E}_{k}}^{H})\!-\!\!\!\!
\sum\nolimits_{\varrho\in\{\text{I},\text{O}\}}\!\!\text{Tr}(\mathbf{X}_{\varrho}\mathbf{W}_{\varrho})\!+\!\iota,
\end{align}
where $\{\lambda_{1},\lambda_{2},\mathbf{X}_{\varrho}\}$ denote the corresponding lagrange multipliers, and $\iota$ is the collection of the terms independent of $\mathbf{W}_{\varrho}$. Checking the KKT conditions, we have
    \begin{subequations}
    \begin{align}
    \label{A-2a}\tag{A-2a} &\text{K1}:\quad \lambda_{1}^{*}\geq 0, \quad \lambda_{2}^{*}\geq 0,,\quad \mathbf{X}_{\varrho}^{*}\succeq \mathbf{0},\\
    \label{A-2b}\tag{A-2b} &\text{K2}:\quad \mathbf{W}_{\varrho}^{*}\mathbf{X}_{\varrho}^{*}=\mathbf{0},\\
    \label{A-2c}\tag{A-2c} &\text{K3}:\quad \nabla_{\mathbf{W}_{\varrho}^{*}}\mathcal{L} \triangleq \mathbf{I}-\mathbf{C}\mathbf{U}_{\text{t}/\text{r}}\mathbf{C}^{H}-\mathbf{X}^{*}=0,
    \end{align}
    \end{subequations}
where $\mathbf{C}=\lambda_{1}^{*}\mathbf{V}_{\varrho}-\lambda_{2}^{*}\mathbf{V}_{\text{E}_{k}}$. With some equivalent algebraic manipulations, it can readily obtain
\begin{align}\tag{A-3}
\mathbf{W}^{*} = \mathbf{C}\mathbf{U}_{\text{t}/\text{r}}\mathbf{C}^{H}\mathbf{W}^{*}.
\end{align}
Hence, we have $\text{rank}(\mathbf{W}^{*})=\text{rank}(\mathbf{C}\mathbf{U}_{\text{t}/\text{r}}\mathbf{C}^{H}\mathbf{W}^{*})\leq \text{rank}(\mathbf{U}_{\text{t}/\text{r}})=\text{rank}(\mathbf{u}_{\text{t}/\text{r}}\mathbf{u}_{\text{t}/\text{r}}^{H})=1$. This completes the proof.


\begin{thebibliography}{99}

\bibitem{M.Di.Renzo_RIS_JSAC}
M. Di Renzo, A. Zappone, et al. ``Smart radio environments empowered by reconfigurable intelligent surfaces: How it works, state of research, and road ahead,'' \textit{IEEE J. Sel. Areas Commun.}, vol. 38, no. 11, pp. 2450--2525, Nov. 2020.

\textit{IEEE Commun. Mag.}, vol. 58, no. 1, pp. 106--112, Jan. 2020.

\bibitem{Yuanwei_IRS_magazine}
Y. Liu, X. Liu, X. Mu, T. Hou, J. Xu, M. Di Renzo, and N. Al-Dhahir, ``Reconfigurable intelligent surfaces: Principles and opportunities,''
\textit{IEEE Commun. Surv. Tut.}, vol. 23, no. 3, pp. 1546--1577, 3rd Quart. 2021.

\bibitem{Lu_IRS_magazine}
Z. Ding, L. Lv, et al. ``A State-of-the-art survey on reconfigurable intelligent surface-assisted non-orthogonal multiple access networks''
\textit{Proc. IEEE.}, to be published, doi: 10.1109/JPROC.2022.3174140.




\bibitem{STAR_magazine}
Y. Liu, X. Mu, J. Xu, R. Schober, Y. Hao, H. V. Poor, and L. Hanzo, ``STAR: Simultaneous transmission and reflection for 360$^{\circ}$ coverage by intelligent surfaces,'' \textit{IEEE Wireless Commun.}, vol. 28, no. 6, pp. 102--109, Dec. 2021.


\bibitem{J.Xu_STAR-RIS}
J. Xu, Y. Liu, X. Mu, and O. A. Dobre, ``STAR-RISs: Simultaneous transmitting and reflecting reconfigurable intelligent surfaces,''
\textit{IEEE Commun. Lett.}, vol. 25, no. 9, pp. 3134--3138, Sep. 2021.

\bibitem{S.Yang_STAR-RIS}
S. Yang, J. Zhang, W. Xia, H. Gao, and H. Zhu, ``Joint power and discrete amplitude allocation for STAR-RIS-aided NOMA system,'' \textit{IEEE Trans. Veh. Technol.}, to be published, doi: 10.1109/TVT.2022.3195815.

\bibitem{H.Niu_STAR-RIS_1}
H. Niu, Z. Chu, F. Zhou, P. Xiao, and N. Al-Dhahir, ``Weighted sum rate optimization for STAR-RIS-assisted MIMO system,'' \textit{IEEE Trans. Veh. Technol.}, vol. 71, no. 2, pp. 2122--2127, Feb. 2022.


\bibitem{X.Mu_STAR-RIS}
X. Mu, Y. Liu, L. Guo, J. Lin, and R. Schober, ``Simultaneously transmitting and reflecting (STAR) RIS aided wireless
communications,'' \textit{IEEE Trans. Wireless Commun.}, vol. 21, no. 5, pp. 3083--3098, May. 2022.


\bibitem{H.Niu_STAR-RIS_security}
H. Niu, Z. Chu, F. Zhou, and Z. Zhu, ``Simultaneous transmission and reflection reconfigurable intelligent surface assisted secrecy MISO networks,''
\textit{IEEE Commun. Lett.}, vol. 25, no. 11, pp. 3498--3502, Nov. 2021.

\bibitem{W.Wang_STAR-RIS_PLS}
W. Wang, W. Ni, H. Tian, Z. Yang, C. Huang, and K. -K. Wong, ``Safeguarding NOMA networks via reconfigurable dual-functional surface under imperfect CSI,'' \textit{IEEE J. Sel. Topics Signal Process.}, to be published, doi: 10.1109/JSTSP.2022.3175013.

\bibitem{Y.Han_STAR-RIS_PLS}
Y. Han, N. Li, Y. Liu, T. Zhang, and X. Tao, ``Artificial noise aided secure NOMA communications in STAR-RIS networks,''
\textit{IEEE Wireless Commun. Lett.}, vol. 11, no. 6, pp. 1191--1195, Jun. 2022.

\bibitem{ZZ_STAR}
Z. Zhang, J. Chen, Y. Liu, Q. Wu, B. He, and L. Yang, ``On the secrecy design of STAR-RIS assisted uplink NOMA networks,'' \textit{IEEE Trans. Wireless Commun.}, to be published, doi: 10.1109/TWC.2022.3190563.


\bibitem{Y.Liu_coupled_STAR}
Y. Liu, X. Mu, R. Schober, and H. V. Poor, ``Simultaneously transmitting and reflecting (STAR)-RISs: A coupled phase-shift model,'' in \textit{Proc. IEEE Int. Conf. Commun. (ICC)}, Seoul, South Korea, May. 2022, pp. 2840--2845.

\bibitem{H.Niu_coupled_STAR}
H. Niu and X. Liang, ``Weighted sum-rate maximization for STAR-RISs-aided networks with coupled phase-shifters,''
\textit{IEEE Syst. J.}, to be published, doi: 10.1109/JSYST.2022.3159551.






\bibitem{T.Jiang_DCP}
T. Jiang and Y. Shi, ``Over-the-air computation via intelligent reflecting surfaces,'' in
\textit{Proc. IEEE Global Commun. Conf. (GLOBECOM)}, Waikoloa, HI, USA, Dec. 2019, pp. 1--6.






\bibitem{PDD-1}
Q. Shi and M. Hong, ``Penalty dual decomposition method for nonsmooth nonconvex optimization--part I: Algorithms
and convergence analysis,'' \textit{IEEE Trans. Signal Process.}, vol. 68, pp. 4108--4122, Jun. 2020.

\end{thebibliography}
\end{document}